\documentclass[useAMS,usenatbib,usegraphicx]{mn2e}
\makeatletter
\def\endfigure{\end@float}
\def\endfigure{\end@float}
\makeatother
\usepackage{afterpage}
\usepackage{float}
\usepackage{subfigure}

\usepackage{amsmath,amssymb}
\usepackage{mathpazo}
\usepackage{bm}
\usepackage{graphicx}
\usepackage{verbatim}
\usepackage{wrapfig}
\usepackage{ascmac}
\usepackage{makeidx}
\usepackage{mathrsfs}
\usepackage{amsfonts}
\usepackage{color}
\def\Vec#1{\mbox{\boldmath $#1$}}
\begin{document}
\title{Statistical properties of filaments in weak gravitational lensing}
\author[Y.Higuchi, M.Oguri and M.Shirasaki]{Yuichi Higuchi$^{1, 2}$\thanks{E-mail: yuichi.higuchi@nao.ac.jp}, Masamune Oguri$^{3, 4}$ and Masato Shirasaki$^{3}$\\
$^{1}$Department of Astronomy, University of Tokyo, 7-3-1 Hongo, Bunkyo-ku, Tokyo 113-0033, Japan\\
$^{2}$Optical and Infrared Astronomy Division, National Astronomical Observatory of Japan, 2-21-1 Osawa, Mitaka, Tokyo 181-8588, Japan\\
$^{3}$Department of Physics, University of Tokyo, 7-3-1 Hongo, Bunkyo-ku, Tokyo 113-0033, Japan\\
$^{4}$Kavli Institute for the Physics and Mathematics of the Universe (Kavli IPMU, WPI), University of Tokyo, Chiba 277-8583, Japan\\}
\maketitle

\begin{abstract}
We study weak lensing properties of filaments that connect clusters of galaxies through large cosmological $N$-body simulations. We select 4639 halo pairs with masses higher than $10^{14}h^{-1}\mathrm{M}_\odot$ from the simulations and investigate  dark matter distributions between two haloes with ray-tracing simulations. In order to classify filament candidates, we estimate convergence profiles and perform profile fitting. We find that matter distributions between haloes can be classified in a plane of fitting parameters, which allow us to select straight filaments from the ray-tracing simulations. We also investigate statistical properties of these filaments, finding them to be consistent with previous studies. We find that $35\%$ of halo pairs possess straight filaments, $4\%$ of which can directly be detected at $S/N\geq2$ with weak lensing. Furthermore, we study statistical properties of haloes at the edges of filaments. We find that haloes are preferentially elongated along filamentary structures and are less massive with increasing filament masses. 
However, the dependence of these halo properties on masses of straight filaments is very weak.
\end{abstract}


\section{Introduction}
The standard model of structure formation that assumes the existence of cold dark matter predicts the hierarchal structure formation (e.g.,  \citealt{1980lssu.book.....P, 2005Natur.435..629S}). As a result, the large scale structure of the Universe shows a complex structure, the so called cosmic web, which is seen in both $N$-body simulations and observations such as the Sloan Digital Sky Survey (SDSS; \citealp{2000AJ....120.1579Y}). Two main components that constitute the cosmic web are voids and filaments. Voids are defined by nearly empty, i.e., underdense region. On the other hand, filaments are slightly overdense regions which have not collapsed into haloes. Massive haloes such as clusters of galaxies form at the intersections of filaments. 

\citet{1996Natur.380..603B} theoretically investigated the formation of filaments as a result of non-linear evolution. Properties of filaments have also been studied through $N$-body simulations \citep{2005MNRAS.359..272C, 2010MNRAS.401.2257M}. Several methods have been developed for characterizing and describing the filamentary structure, including second moment and minimal spanning tree methods \citep{2002SPIE.4847...86M, 2010MNRAS.401.2257M}. In addition, properties of haloes that exist near filaments were studied through $N$-body simulations (e.g., \citealt{2012MNRAS.427.3320C, 2012MNRAS.421L.137L}). These studies showed that the formation processes and  properties of haloes are different between low mass and high mass haloes.

The blind detection of filaments is not easy (e.g., \citealt{1998astro.ph..9268K, 2008MNRAS.383.1655S, 2008MNRAS.385.1431H}). One way to locate filaments is to look for intercluster filaments, i.e., filaments connecting clusters of galaxies. Attempts to detect intercluster filaments with X-ray \citep{1995A&A...302L...9B, 2000ApJ...528L..73S, 2003A&A...403L..29D, 2008PhDT.......221W} have not been very convincing because  it is difficult to distinguish whether those X-ray signals come from filaments or haloes. An alternative way for detecting the distribution of matter in filaments is provided by weak gravitational lensing. Weak lensing is useful because it does not depend on their dynamical state and kind of matter. Indeed there have been several claims of weak lensing detections of filaments between clusters of galaxies \citep{1998astro.ph..9268K, 2002ApJ...568..141G, 2004ogci.conf...34D, 2012Natur.487..202D, 2012MNRAS.426.3369J, 2013ApJ...777...43M}. 

Given developments of methods and several observations for the detection of filaments, more detailed understanding of properties of filaments in simulations are needed. In this paper, we classify dark matter distributions between halo pairs using weak lensing mass maps, and study statistical properties of filaments as well as haloes at the edges of the filaments.
 
This paper is organized as follows. In Section~\ref{sec.analysis}, we describe our analysing techniques, focusing on the basics of weak gravitational lensing and the characterization of haloes. In Section~\ref{sec.simulation}, we describe our simulations, the selection method of filament candidates and the detection method used for searching filaments. In Section~\ref{result}, we describe results of our detection method and properties of haloes. We summarize our results in Section~\ref{conclusion}.

\section{Analysis for filament detection}
\label{sec.analysis}
Let us first summarize weak lensing and characterization for properties of filaments and haloes.
\subsection{Basics of weak lensing}
\label{sec.wl}
 Here we summarize basics of weak lensing for detecting dark matter profile. For more detail description, see \citet{1996MNRAS.283..837S}, \citet{2001PhR...340..291B}, and \citet{2004MNRAS.350..893H}.

Gravitational lensing signals are characterized by the convergence $\kappa(\Vec{\theta})$, and shear $\gamma_1(\Vec{\theta})$ and $\gamma_2(\Vec{\theta})$. The convergence describes the magnification of an image and the shear describes the distortion. While shear is defined relative to a reference Cartesian coordinate frame, it is convenient to consider shear component in rotated frame. In such a frame, tangential and cross shear components $\gamma_+$ and $\gamma_\times$ are defined as
\begin{equation}
\left(
\begin{array}{cc}
\gamma_+   \\
\gamma_\times     
\end{array}
\right)
=
\left(
\begin{array}{cc}
-\mathrm{cos}2\phi&-\mathrm{sin}2\phi \\
-\mathrm{sin}2\phi& \mathrm{cos}2\phi \\
\end{array}
\right)
\left(
\begin{array}{c}
\gamma_1\\
\gamma_2   
\end{array}
\right),
\label{angle}
\end{equation}
where $\phi$ is the angle between axis-$\theta_1$ and $\Vec{\theta}$. 

\begin{figure}
\begin{center}
\includegraphics[width=9cm, bb= 0 0 842 595]{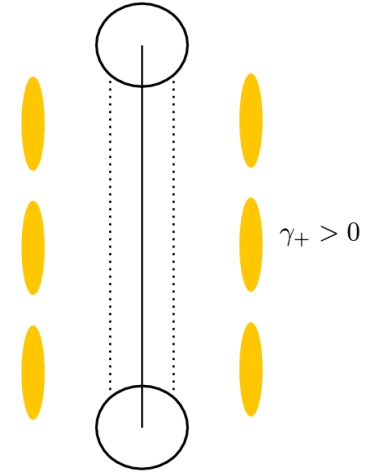}
\caption{Black circles indicate haloes. The halo-halo axis is defined as the line connecting the halo centres (solid line). Dotted line shows a filamentary structure. The shapes of background galaxies are expected to be distorted along the halo-halo axis. The tangential shear which is parallel to the filamentary structure is defined as positive.}
\label{fig.defgamt}
\end{center}
\end{figure}

For a spherical symmetric object, tangential shear is defined relative to the centre of the object. On the other hand, filament is not spherically symmetric but rather axial symmetric. Therefore, in this paper we define "tangential shear" for a filament as a distortion along the filament (Figure~\ref{fig.defgamt}). Specifically, the tangential shear at each point is defined relative to the closest point on the halo-halo axis to the point. With this definition, positive tangential shear ($\gamma_+>0$) describes a distortion parallel to the halo-halo axis. 

Weak lensing signals which are proportional to the density contrast are contaminated by noises, such as intrinsic shapes of galaxies and large scale structure along the line-of-sight. Therefore objects which have a small density contrast such as filaments are difficult to detect with weak lensing. Stacked lensing is a powerful tool to solve this problem and to obtain accurate average profiles because it can reduce those errors \citep{2001PhR...340..291B, 2010PASJ...62..811O, 2011PhRvD..83b3008O, 2012MNRAS.420.3213O, 2013MNRAS.432.1021H}.
 
\subsection{Signal-to-Noise ($S/N$) ratio}
\label{sec.sn}
It is useful to use signal-to-noise ($S/N$) ratio for estimating the observability.
Here we consider shape noise as a source of noise. The shape noise for a bin is obtained as
\begin{equation}
\sigma^2_{\mathrm{noise}, i}=\frac{\sigma^2_{\epsilon}}{2\mathrm{n_gS_i}},
\label{eq.massdis2}
\end{equation}
where $\sigma_{\epsilon}$ is the dispersion of the intrinsic ellipticity distribution, $\mathrm{n_g}$ is a number density of galaxies and $\mathrm{S_i}$ is an area of the bin.
Assuming an axial symmetric filament, the total S/N in the weak lensing is calculated as
\begin{equation}
\left(\frac{S}{N}\right)^2=\frac{2\mathrm{n_g}\mathrm{l}}{\sigma_\epsilon^2}\int\gamma_+^2(\theta)d\theta,
\label{eq.sn2}
\end{equation}
where $\mathrm{l}$ is a length of filament and $\theta$ is a distance from the halo-halo axis. 

Throughout this paper, we adopt $\sigma_{\epsilon}=0.4$ and $\mathrm{n}_\mathrm{g}=30$ arcmin$^{-2}$ which are close to expected values in the Hyper Suprime-Cam (HSC) survey (\citealt{2006SPIE.6269E...9M}). 

\subsection{Model of filament profile}
\label{sec.filpro}
The filament profile has been studied through both simulations and observations \citep{2005MNRAS.359..272C, 2012Natur.487..202D}.
These studies indicate that the convergence profile of filaments is approximately constant inside a scale $\theta_c$ and decline as $\theta^{-2}$ outside. Thus we model the convergence profile of filaments as a function of the distance from the halo-halo axis on the sky plane, $\theta$, as \citep{2005MNRAS.359..272C}
\begin{equation}
\kappa(\theta)=\frac{\kappa_0}{1+\left(\theta/\theta_c\right)^2},
\label{eq.filamentmodel}
\end{equation}
where $\kappa_0$ describes the convergence value at $\theta=0$. We treat $\kappa_0$ and $\theta_c$ as free parameters in the fitting process (Section~\ref{sec:fitting}).

\begin{figure}
\begin{center}
\includegraphics[width=8cm]{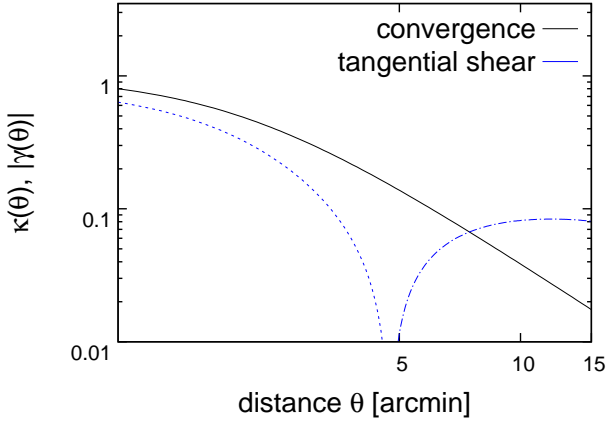}
\caption{Convergence and tangential shear profiles of a filament. The convergence (solid line) and absolute value of tangential shear (dotted and dot-dashed line) are plotted as a function of distance from the filament. Dotted and dot-dashed line show regions in which tangential shear takes negative and positive values, respectively. The tangential shear takes negative values near the centre and positive values at large distances. l$_{2\mathrm{D}}=30$~Mpc, $\kappa_0=1$ and $\theta_c=2$~arcmin are assumed.}  
\label{fig:anamodel}
\end{center}
\end{figure}

We compute the corresponding tangential shear as follows. From the convolution theorem, the Fourier transform of complex shear $\hat{\gamma}$ is obtained with two dimensional wave vector $\Vec{l}$ as \citep{2001PhR...340..291B}
\begin{equation}
\hat{\gamma}(\Vec{l})=\frac{1}{\pi}\hat{\mathscr{D}}(\Vec{l})\hat{\kappa}(\Vec{l}),
\label{eq.ftgam}
\end{equation}
where $\hat{\mathscr{D}}(\Vec{l})$ is the Fourier transform of the complex kernel $\mathscr{D}$ defined as
\begin{equation}
\mathscr{D}(\Vec{\theta})=\frac{-1}{(\theta_1-\mathrm{i}\theta_2)^2}.
\end{equation}
Assuming the convergence profile with equation~(\ref{eq.filamentmodel}) and a length of the filament l$_{2\mathrm{D}}$ along $\theta_2$-axis, the Fourier transform of convergence is calculated as
\begin{eqnarray}
\hat{\kappa}(\Vec{l})&=&\int^\infty_{-\infty}d\theta_1\int^\infty_{-\infty}d\theta_2\kappa(\Vec{\theta})\mathrm{exp}\left(\mathrm{i}\Vec{\theta}\cdot\Vec{l}\right)\Theta\left(\frac{\mathrm{l_{2D}}}{2}-\left|\theta_2\right|\right)\nonumber\\
&=&\frac{2\pi\theta_c\kappa_0}{l_2}\left\{\mathrm{exp}\left(l_1\theta_c\right)\Theta(-l_1)\right.\nonumber\\ 
        &&\left.+\mathrm{exp}\left(-l_1\theta_c\right)\Theta(l_1)\right\}\mathrm{sin}\left(\frac{\mathrm{l_{2D}}}{2}l_2\right)\Theta(\mathrm{l_{2D}}),
\label{eq.ftcon}
\end{eqnarray}
where $\Theta$ is the Heaviside step function. Then the tangential shear is calculated by carrying out the inverse Fourier transformation (equation~\ref{eq.ftgam}). Figure~\ref{fig:anamodel} shows convergence and tangential shear profiles. We assume the parameter set l$_{2\mathrm{D}}=30$ Mpc, $\kappa_0=1$ and $\theta_c=2$ arcmin. Tangential shear takes negative values near the centre of the filament and positive values at large distances.

\subsection{Properties of haloes}
\label{sec.halo}
 In this section, we introduce halo properties. For more details, see \citet{2004MNRAS.350..893H, 2012MNRAS.425.2287H}.

 The density profile of dark matter haloes is described by the NFW profile \citep{1997ApJ...490..493N} which is given by 
\begin{equation}
\rho_{\mathrm{NFW}}\left(\mathrm{r}\right)=\frac{\rho_\mathrm{s}}{\left(\mathrm{r/r_s}\right)(1+\mathrm{r/r_s})^2},
\label{eq.NFW}
\end{equation}
where $\rho_\mathrm{s}$ is the density parameter and $\mathrm{r_s}$ is the scale radius. 
The halo mass $\mathrm{M}_\Delta$, the density of which is $\Delta$ times the critical density, is defined as
\begin{equation}
\mathrm{M}_\Delta=\frac{4\pi}{3}\Delta\rho_\mathrm{cr}\mathrm{r}_\Delta^3.
\end{equation}  

To study the connection between shapes of haloes and properties of filaments, we also consider a triaxial model \citep{2002ApJ...574..538J}. In this model, equation~(\ref{eq.NFW}) is modified as
\begin{equation}
\rho_\mathrm{tri}=\frac{\rho_{\mathrm{cd}}}{(\mathrm{R/R_0})(1+\mathrm{R/R_0})^2},
\label{eq.triaxial1}
\end{equation}
\begin{equation}
\mathrm{R}^2=\mathrm{c}^2\left(\frac{x^2}{\mathrm{a}^2}+\frac{y^2}{\mathrm{b}^2}+\frac{z^2}{\mathrm{c}^2}\right)\hspace{1cm}(\mathrm{a\leq b\leq c}). 
\label{eq.triaxial2}
\end{equation}
The axial ratios $\mathrm{a/b}$ and $\mathrm{a/c}$ are used for characterizing the shapes of haloes. 
In practice, the axis ratios are computed via the inertia tensor of the haloes defined by
\begin{equation}
\mathrm{I}_{\mathrm{ij}}=\int d\Vec{x}(x_\mathrm{i}-\bar{x}_\mathrm{i})(x_\mathrm{j}-\bar{x}_\mathrm{j})\rho(\Vec{x}),
\label{eq.inertia}
\end{equation}
where summation indices i, j run from 1 to 3 \citep{2012MNRAS.425.2287H}. We can evaluate inertia tensors along an arbitrary direction by using diagonal and non-diagonal elements of the inertia tensor. In Section~\ref{result}, inertia tensors $\mathrm{I_{rr}}$ and $\mathrm{I}_{\theta\theta}$ which are along a halo-halo axis and a perpendicular direction of it are used to study the correlation between orientation of haloes and filament directions. In order to distinguish mass distributions between haloes and filaments, the halo properties are calculated within the region which exceed 500 times the critical density.

\section{Simulation}
\label{sec.simulation}
\subsection{Simulation data}
\label{simulationdata}

We use a large set of ray-tracing simulations in $N$-body simulations (see \citealt{2009ApJ...701..945S, 2011MNRAS.414.1851O, 2012MNRAS.425.2287H, 2013MNRAS.432.1021H}). The $N$-body simulations were performed with $256^3$ particles in a cubic box of $240h^{-1}$ Mpc  on a side with the code Gadget-2. Cosmological parameters according to the WMAP 3-year result \citep{2007ApJS..170..377S} were assumed; Hubble parameter $\mathrm{H_0}=73.2$ km/s/Mpc, matter density $\Omega_\mathrm{m}=0.238$, baryon density $\Omega_\mathrm{b}=0.042$, dark energy $\Omega_\Lambda=0.762$, dark energy equation of state parameter $\mathrm
{w}=-1$, spectral index $n_s=0.958$ and variance of density fluctuation within $8h^{-1}$~Mpc $\sigma_8=0.76$. Ray-tracing is performed following the method described in \citet{2001MNRAS.327..169H}.
In the simulation, source redshift is fixed to $\mathrm{z_s}=0.997$. Field of view (FOV) for each realization is $5\times5$ degree$^2$ with angular grid size $\sim0.15$ arcmin and with $200$ realizations the total FOV is $5000$ degree$^2$. The convergence power spectrum is consistent with the result in the high resolution simulation down to angular wavenumber $l\lesssim10^4$ at $5\%$ level (\citealt{2009ApJ...701..945S}). This result assures the accuracy for our analysis down to arcmin scale.

The value of matter density parameter in the simulations is lower than the result estimated from Planck (\citealt{2013arXiv1303.5076P}). The difference may change the number counts of filaments and the statistical properties of haloes connecting the filaments.
\subsection{Candidate of filaments}
\label{sec.canfil}
\begin{figure}
\begin{center}
\includegraphics[width=9cm]{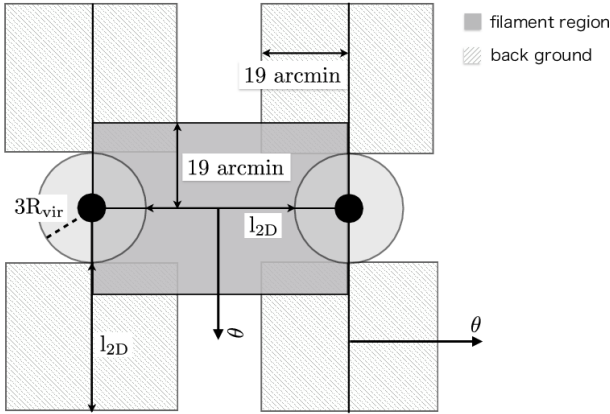}
\caption{Selection of halo pairs for analysing filaments. Black circles are haloes. Regions inside $19$ arcmin on the sky plane, which corresponds to $5h^{-1}$ Mpc on the lens plane, from the halo-halo axis and the axes which are perpendicular to the halo-halo axis are shaded. If haloes with masses lager than $10^{14}h^{-1}\mathrm{M}_\odot$ exist in the shaded regions, those halo pairs are excluded. The convergence profiles in a filament region and background regions are estimated as a function of distance from the axes we defined. In the process of estimating the filament profile, we exclude the region within $3\mathrm{R}_\mathrm{vir}$ from the centres of haloes in order to reduce the contribution from haloes inside that region.}
\label{fig.exregion}
\end{center}
\end{figure}

 Here we summarize the method to select halo pairs for carrying out profile fitting (Section~\ref{sec:fitting}). Figure~\ref{fig.exregion} illustrates the procedure which we explain in this section. 

 First, haloes with Friend-of-Friend (FOF) masses $\mathrm{M}\geq10^{14}h^{-1}\mathrm{M}_\odot$ and in the redshift range $0.4\leq \mathrm{z}\leq0.6$ are extracted from the halo catalogue. 
We select halo pairs with separation smaller than $60$ Mpc. Then for each halo pair we define the halo-halo axis which connects both haloes. We are interested in properties of filamentary structures. In order to eliminate contaminations of other haloes along the line-of-sight, halo pairs are excluded if any haloes with mass $\mathrm{M}\geq10^{14}\mathrm{M}_\odot$ along the line-of-sight lie within $\sim19$ arcmin on the sky plane from the halo-halo axis, which corresponds to $5h^{-1}$ Mpc on the lens plane. 
 
 Next, in order to fit the convergence profile with equation~(\ref{eq.filamentmodel}), we estimate the contribution from the two haloes and subtract it from the profile. The contribution is estimated from four new axes which are perpendicular to the halo-halo axis (see Figure~\ref{fig.exregion}). The lengths of these new axes are the same as that of the halo-halo axis. Then halo pairs are also excluded if other haloes with mass $\mathrm{M}\geq 10^{14}h^{-1}\mathrm{M}_\odot$ exist within $19$ arcmin from these axes. 
 
As a result of this selection, the total number of halo pairs for examining filament structures is $4639$. 

\subsection{Profile fitting}
\label{sec:fitting}
In order to classify filament candidates, we perform profile fitting of background subtracted convergence profiles. In this section, we summarize the fitting process for filament candidates.

The value of tangential shear is not a local value, i.e., affected by other structures such as haloes exist at both ends of filaments. On the other hand, the value of convergence is a local value, i.e., trace the distribution of matters at local points. In order to avoid contaminations from neighboring structure, fitting is performed in convergence map. Fitting is carried out to convergence profiles without shape noise.

Since we want to detect weak lensing signals from filaments, the contribution from haloes must be excluded.  We do not use the region within $3\mathrm{R_\mathrm{vir}}$ from the centres of both haloes since the infall region of a cluster extends out to about three times its virial radius \citep{1997ApJ...481..633D}.  Therefore the projected distance $\mathrm{L_{2D}}$ between two haloes is related with the virial radii and projected filament length $\mathrm{l_{2D}}$ as
\begin{equation}
\mathrm{L_{2D}}=3(\mathrm{R}_{\mathrm{vir},1}+\mathrm{R}_{\mathrm{vir}, 2})+\mathrm{l_{2D}},
\end{equation}
where $\mathrm{R}_{\mathrm{vir}, i}$ is a virial radius for each halo. In this paper, we use $\mathrm{L}$ for the distance between two haloes and $\mathrm{l}$ for the filament length. The convergence profile in a filament region is estimated for the remaining region as a function of a distance from halo-halo axis. While we exclude the region near haloes to reduce halo contributions,  the contribution from a diffuse component of haloes, which is named background, still remains. Therefore it should be removed from the estimated profiles of filaments. Because haloes may be elongated in the direction of the halo-halo axis owing to the existence of a filament, it is difficult to estimate the background from the halo profile on the opposite sides of the filament. Thus, the contribution is estimated by using the regions which are perpendicular to the halo-halo axis (Figure~\ref{fig.exregion}). In estimating the background, we impose the same exclusion condition which are adopted to the filament region on these region, i.e., exclude the region within $3\mathrm{R}_\mathrm{vir}$ from the haloes.
When carrying out fitting, the background level is subtracted from the convergence profile estimated in the filament region. We however note that typical background level is much lower than filament signals. 

We treat $\kappa_0$ and $\theta_c$ in equation~(\ref{eq.filamentmodel}) as free parameters to fit the background-subtructed profile. 
The model parameters are determined by minimizing the value of chi-square defined as
\begin{equation}
\chi^2=\sum_{i}\frac{\left[\kappa(\theta_i)-\kappa_\mathrm{model}(\theta_i| \kappa_0, \theta_c)\right]^2}{\sigma_i^2},
\end{equation}
where $\kappa(\theta_i)$ and $\kappa_\mathrm{model}(\theta_i| \kappa_0, \theta_c)$ are convergence in the simulations and  the model at $i$th angular bin on the sky plane. $\sigma_i$ is the error at $i$th bin, which is defined as
\begin{equation}
\sigma_i^2=\sum^{\mathrm{N}_i}_{j=1}\frac{\left[\kappa_j(\theta_i)-\langle\kappa(\theta_i)\rangle\right]^2}{\mathrm{N}_i},
\end{equation}
where $\langle\kappa(\theta_i)\rangle$ and $\mathrm{N}_i$ are average convergence value and number of data in $i$th bin, respectively. $\kappa_j(\theta_i)$ is $j$th pixel value of convergence in $i$th bin.

In order to calculate a convergence profile up to $15$ arcmin from the halo-halo axis, we divide the distance into $10$ bins with logarithmic scale. However, we have not confirm how the choice of bin size affects our fitting result.  
Parameters are searched by using the down hill simplex method, generating random numbers as initial condition. We have confirmed that the choice of the initial condition does not affect the result of our analysis. Fitting is performed in the parameter range $\kappa_0\geq10^{-5}, \theta_c\geq10^{-3}$.

\section{Result}
\label{result}
\subsection{Classification of filaments}
Profile fitting with equation~(\ref{eq.filamentmodel}) was carried out for filament candidates selected in Section~\ref{sec.canfil}. We treat $\kappa_0$ and $\theta_c$ as free parameters. Fitting for each filament candidate is performed to the estimated filament profiles within $15$ arcmin from the halo-halo axis.
 
\begin{figure*}
\subfigure{\includegraphics[width=0.92\columnwidth]{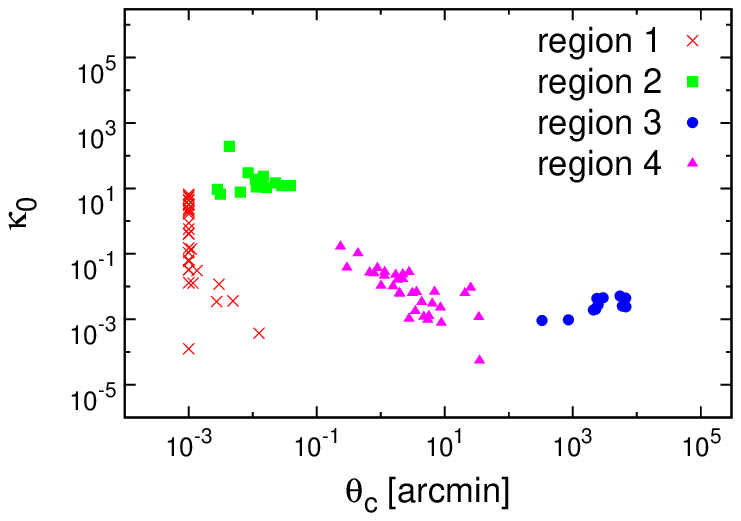}\label{k0-tc-1}}\hspace{1cm}
\subfigure{\includegraphics[width=0.92\columnwidth]{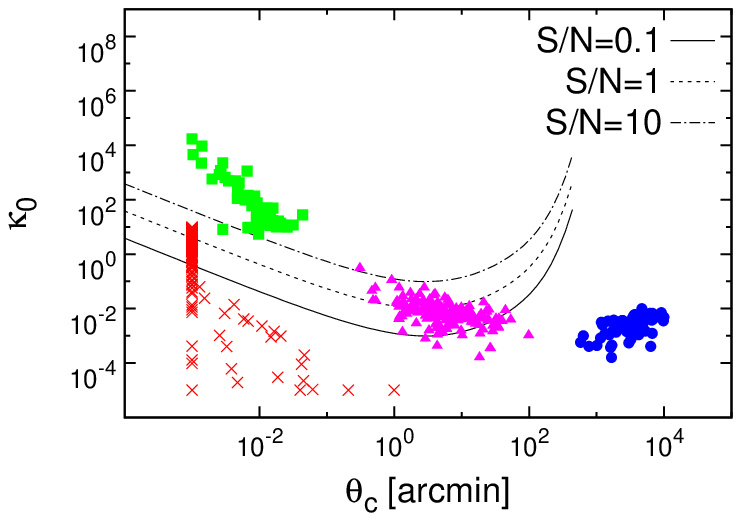}\label{k0-tc-2}}\hspace{1cm}
\subfigure{\includegraphics[width=0.92\columnwidth]{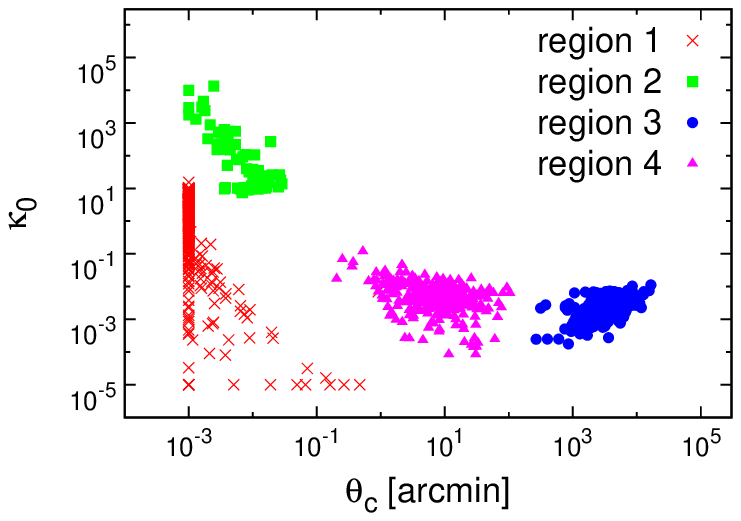}\label{k0-tc-3}}\hspace{1cm}
\caption{Fitting result in the $\kappa_0-\theta_c$ plane. Fitting is carried out for the filament candidates in the redshift range $0.4\leq \mathrm{z}\leq0.6$. The space is divided into 4 regions. Crosses, squares, circles and triangles points indicate region~1, region~2, region~3 and region~4. The parameter sets ($\kappa_0$, $\theta_c$) for each region take ($0.00001\leq\kappa_0\lesssim50$, $0.001\leq\theta_c\lesssim$1), ($50\lesssim\kappa_0$, $0.001\lesssim\theta_c$), ($0.00001\leq\kappa_0<\infty$, $200\lesssim\theta_c$) and ($0.00005\lesssim\kappa_0\lesssim1$, $0.1\lesssim\theta_c\lesssim200$). The fitting results are plotted by their three dimensional distance between two haloes {\it upper left}: $0\leq \mathrm{L_{3D}}\leq20$ Mpc; {\it upper right}: $20\leq \mathrm{L_{3D}}\leq40$ Mpc and {\it lower}: $40\leq \mathrm{L_{3D}}\leq60$ Mpc. Lines in the upper right panel indicate weak lensing detectability with $S/N=0.1$ ({\it solid line}), $S/N=1$ ({\it dotted line}), $S/N=10$ ({\it dot-dashed line}), respectively.}
\label{fig.k0-tc}
\end{figure*}

We divide filament candidates by their three dimensional length ($0\leq \mathrm{L_{3D}}\leq20$ Mpc, $20\leq \mathrm{L_{3D}}\leq40$ Mpc and $40\leq \mathrm{L_{3D}}\leq60$ Mpc). Figure~\ref{fig.k0-tc} shows the result of the fitting in the $\kappa_0-\theta_c$ plane for each subsample. We find that in this plane the filament candidates can be divided into 4 regions defined as region~1 ($0.00001\leq\kappa_0\lesssim50$ and $0.001\leq\theta_c\lesssim1$), region~2 ($50\lesssim\kappa_0$ and $0.001\lesssim\theta_c$), region~3 ($0.00001\leq\kappa_0$, $200\lesssim\theta_c$) and region~4 ($0.00005\lesssim\kappa_0\lesssim1$ and $0.1\lesssim\theta_c\lesssim200$). These criterion are determined by visual inspection. This trend does not depend on the distances between the two haloes.

\begin{figure*}
\begin{center}
\vspace{2cm}
\subfigure{\includegraphics[width=0.52\columnwidth]{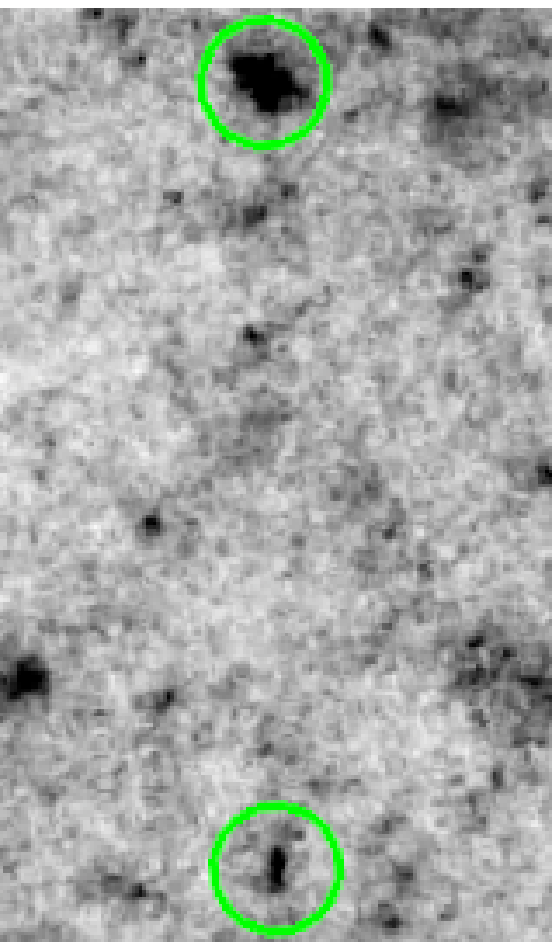}\label{fig.region1-im}}\hspace{1cm}
\subfigure{\includegraphics[width=1.1\columnwidth]{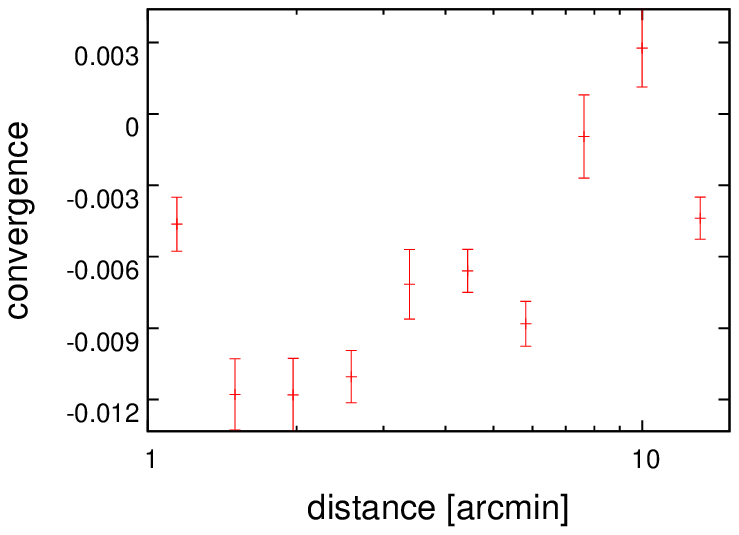}\label{fig.region1-con}}\hspace{1cm}
\end{center}
\caption{An example of filaments in region~1. {\it Left}: convergence map of the filament (linear scale). Circles show the haloes that define the filament candidate. There is no filamentary structure between the haloes. {\it Right}: convergence profile. X-axis is distance from the halo-halo axis. Y-axis is the value of convergence. The points with error bar show average values of convergence in each bin. The values of convergence is computed from convergence map without shape noise} 
\label{fig.region1}
\end{figure*}

{\bf Region~1}: In this region, both parameters are quite small. This suggests that there is no filamentary structure for these halo pairs. Figure~\ref{fig.region1} shows an example of a convergence map and filament profile in this region. Indeed there are not any signals between two haloes in the convergence map.

\begin{figure*}
\begin{center}
\subfigure{\includegraphics[width=0.52\columnwidth]{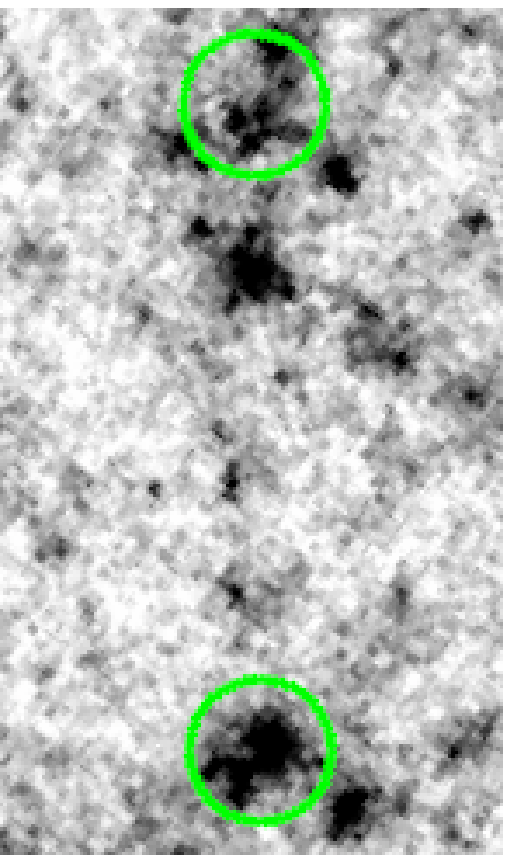}\label{fig.region2-im}}\hspace{1cm}
\subfigure{\includegraphics[width=1.1\columnwidth]{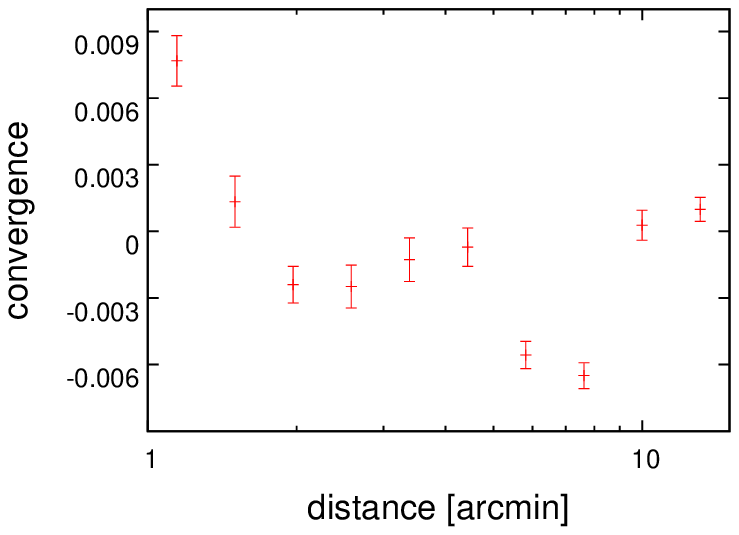}\label{fig.region2-con}}\hspace{1cm}
\end{center}
\caption{Similar to Figure~\ref{fig.region1}, but an example in region~2 is shown.} 
\label{fig.region2}
\end{figure*}

{\bf Region~2}: The values of $\kappa_0$ are large. On the other hand, the scale lengths $\theta_c$ are as small as region~1. The reason why these parameters take such values is existence of small compact structures such as haloes. As shown in Figure~\ref{fig.region2}, there are lower-mass haloes between the two haloes. The small values of the scale length are almost consistent with their virial radii.
Therefore we conclude that there is no filamentary structure in this region.

\begin{figure*}
\begin{center}
\vspace{2cm}
\subfigure{\includegraphics[width=0.52\columnwidth]{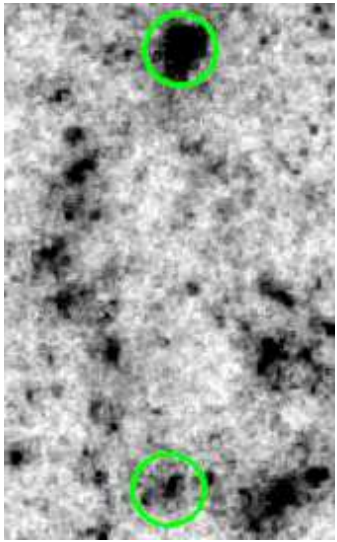}\label{fig.region3-im-1}}\hspace{1cm}
\subfigure{\includegraphics[width=1.1\columnwidth]{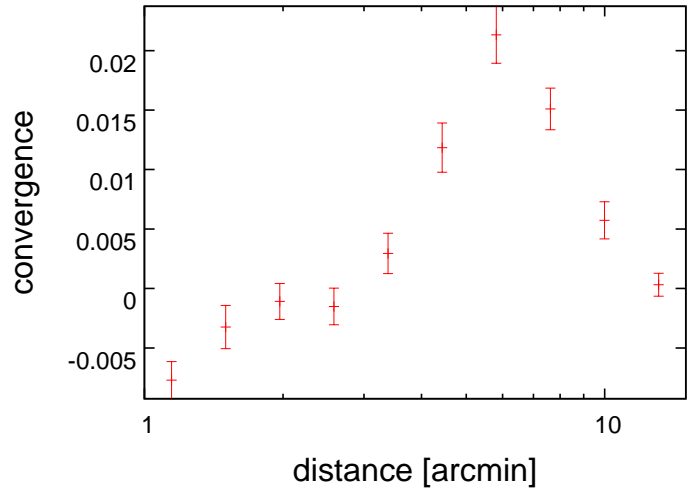}\label{fig.region3-con-1}}\hspace{1cm}
\subfigure{\includegraphics[width=0.52\columnwidth]{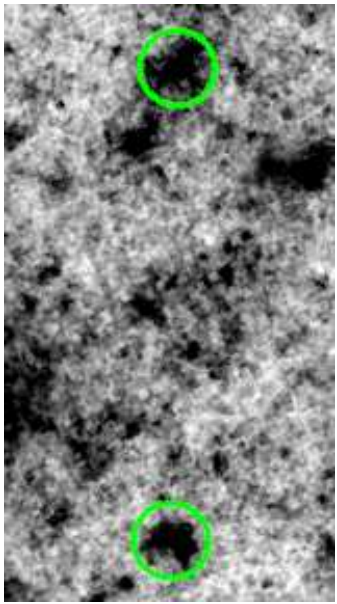}\label{fig.region3-im-2}}\hspace{1cm}
\subfigure{\includegraphics[width=1.1\columnwidth]{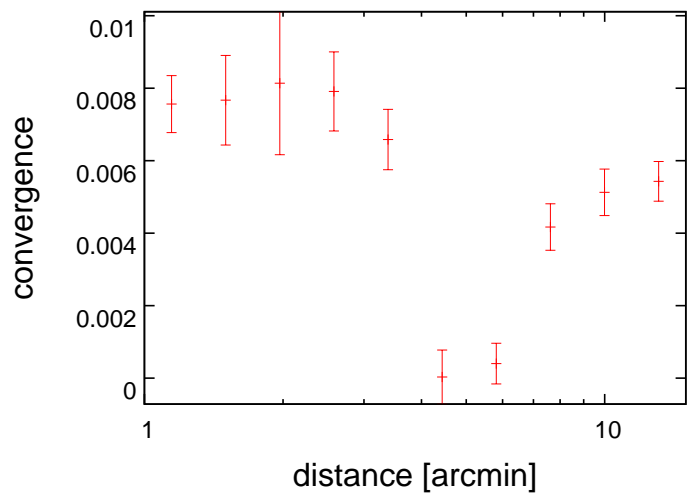}\label{fig.region3-con-2}}\hspace{1cm}
\end{center}
\caption{Similar to Figure~\ref{fig.region1}, but examples in region~3 are shown.} 
\label{fig.region3}
\end{figure*}

{\bf Region~3}: The values of the scale length parameter are quite large in this region, because of the existence of clusters or curved filaments. Figure~\ref{fig.region3} indicate that the convergence profiles and convergence maps have broad or offset profiles. 
Fitting this filament profile with equation~(\ref{eq.filamentmodel}) requires unusually large values of the scale length parameter $\theta_c$.

\begin{figure*}
\begin{center}
\vspace{2cm}
\subfigure{\includegraphics[width=0.52\columnwidth]{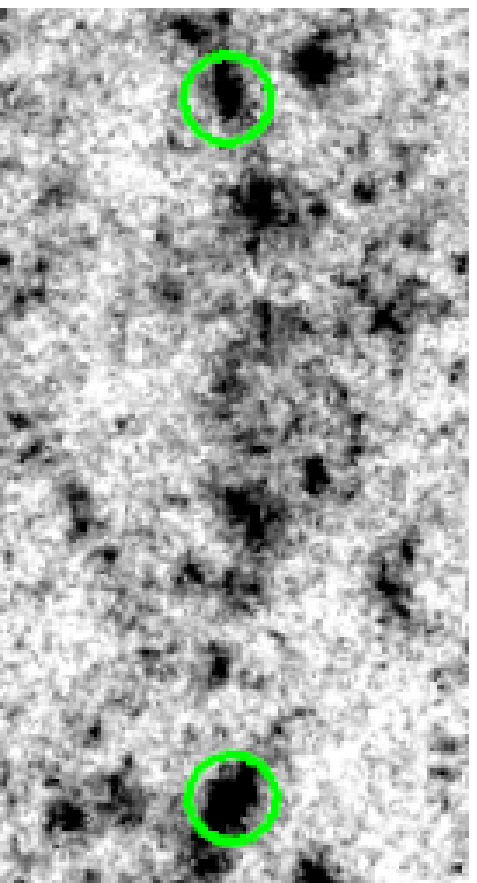}\label{fig.region4-im}}\hspace{1cm}
\subfigure{\includegraphics[width=1.1\columnwidth]{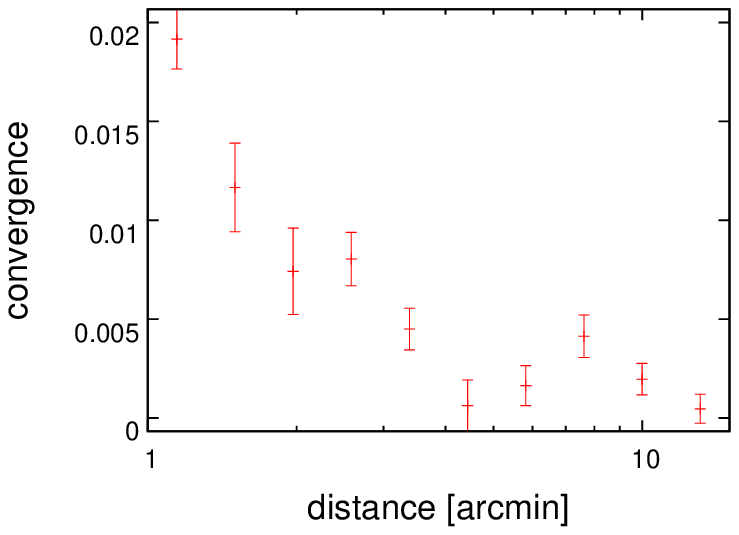}\label{fig.region4-con}}\hspace{1cm}
\end{center}
\caption{Similar to Figure~\ref{fig.region1}, but an example in region~4 is shown.}
\label{fig.region4}
\end{figure*}

{\bf Region~4}: As shown in Figure~\ref{fig.region4}, there is a straight filamentary structure between two haloes. The values of $\theta_c$ are consistent with observed and simulation results (e.g., \citealt{2005MNRAS.359..272C, 2010MNRAS.401.2257M, 2012Natur.487..202D}). Therefore filament candidates which exist in this region are regarded to have straight filamentary structures.

\begin{figure*}
\subfigure{\includegraphics[width=0.92\columnwidth]{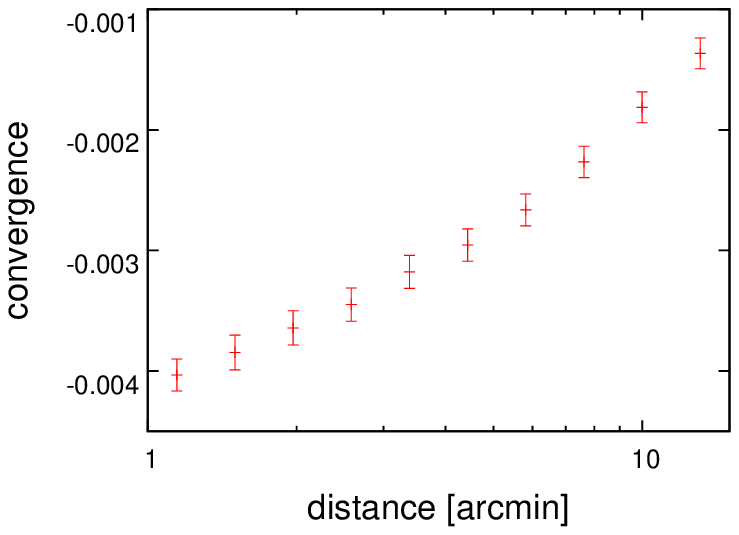}}\hspace{1cm}
\subfigure{\includegraphics[width=0.92\columnwidth]{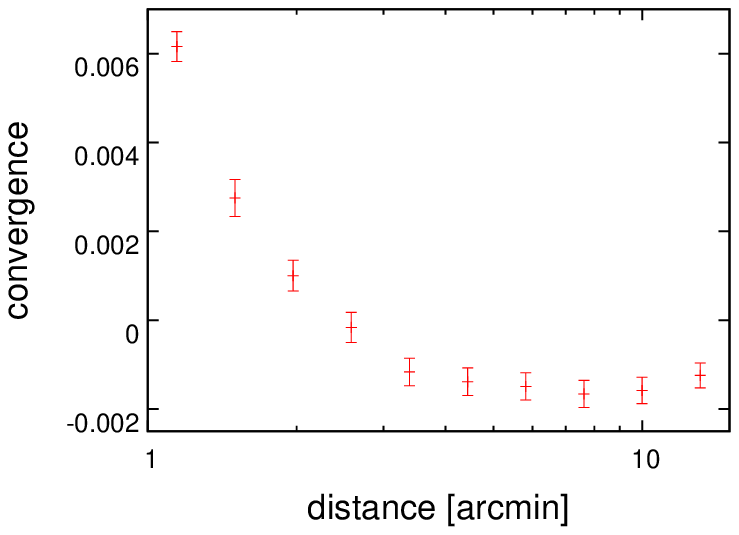}}\hspace{1cm}
\subfigure{\includegraphics[width=0.92\columnwidth]{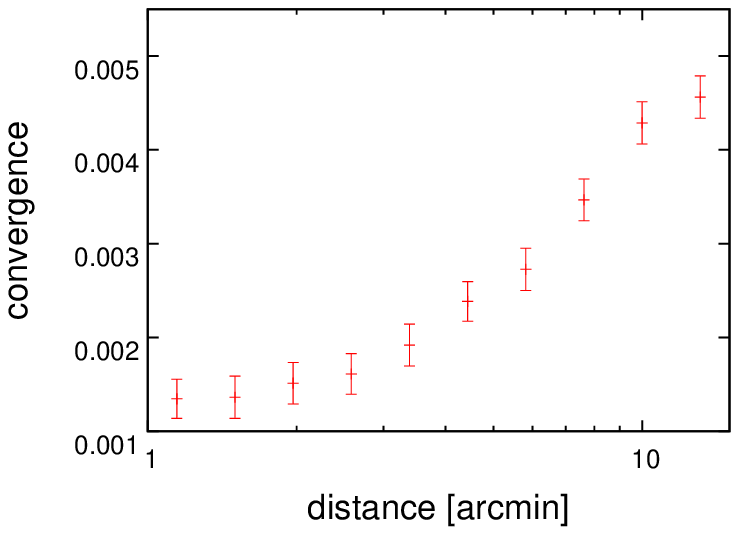}}\hspace{1cm}
\subfigure{\includegraphics[width=0.92\columnwidth]{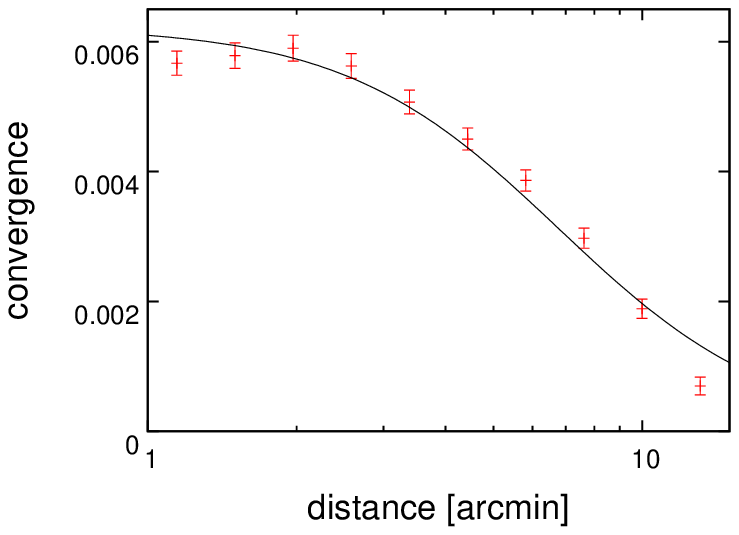}}
\caption{Stacked filament profile for each region. X-axis shows distance from the halo-halo axis. Y-axis shows convergence. {\it upper left}: region~1; {\it upper right}: region~2; {\it lower left}: region~3; {\it lower right}: region~4. The solid line in the lower right panel shows a best fit line with equation~(\ref{eq.filamentmodel}).}
\label{fig.stackedpro}
\end{figure*}

Figure~\ref{fig.stackedpro} shows stacked convergence profiles for all regions. Stacking is carried out for all filament candidates after the background profiles are subtracted. For region~1, the average value of convergence is negative, i.e., underdense region. The values of profile in region~3 increase as the distance increases because of the contributions from curved filaments or offsetted nearby haloes. Compared with the profile in region~4, the profile in region~2 is a much steeper, because the profile is dominated by single haloes.
Stacked lensing profiles also support our visual classification. For a subsample of region~1, the values of convergence estimated from the simulation are $\kappa(\theta)\leq0$ over the considered scale. To reproduce these profiles with equation~(\ref{eq.filamentmodel}), $\kappa_0$ or $\theta_c$ has to take small values. In region~2, $\kappa_0$ sometimes takes too large values, which is partly due to the lack of resolution in our profile fitting. If $\theta_c$ is much smaller than the bin width of profile fitting, $\kappa_0$ is determined by the extrapolation from the outer profile. In region~3, the stacked profile has a off-centred profile due to off-centred mass distributions of individual filaments. In order to minimize chi-square values,  these off-centred profiles cause large values in $\theta_c$. 

\begin{figure}
\begin{center}
\includegraphics[width=9cm]{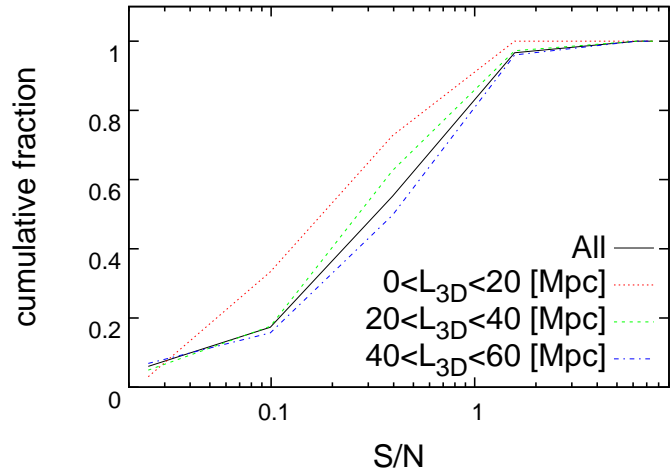}
\end{center}
\caption{Cumulative fraction of filaments as a function of  $S/N$ for weak lensing detection. $S/N$ is computed with estimated parameters $\kappa_0$ and $\theta_c$ through equation~(\ref{eq.sn2}). Solid, dotted, dashed and dot-dashed lines show the fractions for all samples, samples which have a length of $0<\mathrm{L_{3D}}<20$ Mpc, $20<\mathrm{L_{3D}}<40$ Mpc, $40<\mathrm{L_{3D}}<60$ Mpc, respectively. }  
\label{fig.cum}
\end{figure}

 In estimating the detectability with weak lensing, we assume a simple filament model that the filament has a length $\mathrm{l_{2D}}$ of $30$ Mpc and follows the profile defined in equation~(\ref{eq.filamentmodel}).
 Parameters which are related to weak lensing follow those defined in Section~\ref{sec.sn}. In Figure~\ref{fig.k0-tc}, we show the detection significance of $S/N=0.1$, $S/N=1$, $S/N=10$. $S/N$ is computed with parameters estimated through equation~(\ref{eq.sn2}). We find that filaments in region~2 and 4 are detectable. Most of filaments in region~2 can be detected at significant level. On the other hand, The majority of filaments in the region~4 are difficult to detect individually. Figure~\ref{fig.cum} shows the cumulative fraction of filaments as a function of $S/N$ for region~4. Here the filament length $\mathrm{l}_{2\mathrm{D}}$ is estimated from information of halo catalogue for each filament candidate. We find that only a few percent of filaments in region~4 can be detected at $S/N\geq1$. 


\subsection{Statistical properties}
\label {statistics}
In this section, we discuss statistical properties of filaments and haloes classified by our method.
  
 In order to characterize properties of haloes, we define the inertia moment $\mathrm{I_{rr}}$ along a halo-halo axis and $\mathrm{I}_{\theta\theta}$ along a perpendicular direction to that axis, which are obtained by transforming the inertia moments estimated with equation~(\ref{eq.inertia}). We also compute total mass $\mathrm{M}$ of filaments by 
\begin{eqnarray}
\hspace{2cm}\mathrm{M}&=&\mathrm{l}_{2\mathrm{D}}\int_0^{\theta_\mathrm{out}}d\theta\frac{\kappa_0}{1+\left(\theta/\theta_c\right)^2}\nonumber\\ 
&=&\kappa_0\mathrm{l}_{2\mathrm{D}}\theta_c \mathrm{arctan}\left[\frac{\theta_\mathrm{out}}{\theta_c}\right].
\label{eq.mass}
\end{eqnarray}
We adopt $\theta_\mathrm{out}=15$ arcmin and estimated fitting parameters for computing mass of each filament.
 
\subsubsection{Statistics of filaments}
\label{st.fil}
 From our analysis, filament candidates which exist in region~3 and 4 have filaments. Since the fraction of curved filaments is $\sim1.6\%$ of halo pairs \citep{2005MNRAS.359..272C}, we assume that only region~4 have filament. Then $35\%$ filament candidates exist in this region (Table~\ref{tab.fraction1}). This value is more than twice as large as the previous result of \citet{2005MNRAS.359..272C}. There are some explanations for this discrepancy. For example, due to the projection effect we can misidentify structures along the line-of-sight as filaments. It was estimated that about $15\%$ of filament candidates are affected by this projection effect (\citealt{2005MNRAS.359..272C}). Once the estimated value is subtracted from our result, the fraction is consistent with \citet{2005MNRAS.359..272C} at a few percent level. Another reason is the effect of profile fitting. In order to determine two parameters, the average value of convergence is estimated as a function of distance from the halo-halo axis. Therefore if some massive haloes exist in the filament region like the situation in region~2, our method classifies these candidates as a member of region~4. In addition, the difference in the definition of filaments may also explain this discrepancy.
 
 In Table~\ref{tab.fraction2}, we show the fraction of filaments in each region as a function of separation of haloes. We find that the fractions in region~1 and region~3 increase as the separations between haloes increase. 
 For region~3, since the line-of-sight projection increases with increasing separation, the fraction should also increase as a function of separation. In region~2 and region~4, on the other hand, we find the decreasing trend as the separation increases. Since haloes on line-of-sight increase with increasing separation and are randomly distributed between two haloes, the line-of-sight effect is more pronounced in region~3 than in region~2 for large halo separations. Region 4 has a decreasing trend except the case $0\leq \mathrm{L_{3D}}\leq20$ Mpc. This trends is a consistent result studied in \citet{2005MNRAS.359..272C}. Since the small number of samples causes a large statistical error, we do not find the trend for the case $0\leq \mathrm{L_{3D}}\leq20$ Mpc. We expect that the trend is confirmed in future work using larger simulations.
 
 We showed that some filaments in region~4 can be detected at a significant level with weak gravitational lensing. In our simulation, we assume $\mathrm{z_s}=0.996$, $\mathrm{n}_{\mathrm{g}, \mathrm{sim}}=30$ arcmin$^{-2}$ and FOV=$5000$ degree$^2$. In the actual HSC survey, the parameters become as FOV$=1400$ degree$^2$ and $\mathrm{n}_{\mathrm{g}, \mathrm{HSC}}=24$ arcmin$^{-2}$, where only galaxies that exist at $\mathrm{z}>0.6$ are assumed to be used. Therefore signal-to-noise ratio $(S/N)_\mathrm{sim}$ estimated in our simulation are degraded by a factor of about $\sqrt{4/5}$ for individual filaments. Since the number of filament in region 4 is $1698$ and $4.25\%$ of filaments in region~4 are detected at $(S/N)_\mathrm{sim}=(S/N)_\mathrm{HSC}/\sqrt{4/5}\geq2.26$ in our simulation, about 21 filaments will be detected at $(S/N)_\mathrm{HSC}\geq2$ in the entire HSC survey footprint. Therefore the detection of individual filamentary structures through weak gravitational lensing as \citet{2012Natur.487..202D} is rare and challenging. By using the stacked lensing method, however, we can significantly detect weak lensing signals from filaments. The average $S/N$ of filaments in region~4 is $\sim1.1$, and thus by combining 21 filaments we can achieve significant weak lensing detection at $S/N=5$, if we can identify filaments in region~4 from, say, galaxy distributions. Figure~\ref{fig.cum} suggests that stacked weak lensing signal may be dominated by a few massive filaments. Even for typical filaments with $S/N\sim0.3$, significant ($S/N>5$) detection can be achieved by stacking $\sim300$ region~4 filaments, which require $<1000$ degree$^2$ survey area.

\begin{table*}
\begin{center}
\caption{Details of each region. Column (1): region name; Column (2): fraction of the number in each region to the total number of filament candidates; Column(3): classification for the filament candidates in each region.; Column~(4): Fraction estimated in \citet{2005MNRAS.359..272C}.}
\begin{tabular}{cccc}
\hline
&fraction [$\%$]&classification&fraction [\%] in \citet{2005MNRAS.359..272C}\\ \hline\hline
region 1 &37.6&no matter distribution&79\\ 
region 2 &9.26&existence of small halos&10.1$\sim$0\\ 
region 3 &17.8&curved filament or broad mass distribution&3.7\\ 
region 4 &35.3&straight filament&7.2$\sim$17.3\\
\hline
\label{tab.fraction1}
\end{tabular}
\end{center}
\end{table*}
\begin{table*}
\begin{center}
\caption{Fractions of each region for each separation of haloes. Column (1): region name; Column (2), Column (3) and Column (4): fraction of the number in each region to the total number of filament candidates whose haloes have the length of $0\leq \mathrm{L_{3D}}\leq20$ Mpc, $20\leq \mathrm{L_{3D}}\leq40$ Mpc and $40\leq \mathrm{L_{3D}}\leq60$ Mpc for each. Values in parentheses indicate a $1\sigma$ poisson error.}
\begin{tabular}{cccc}
\hline
&fraction ($1\sigma$) [$\%$]&fraction ($1\sigma$) [$\%$]&fraction ($1\sigma$) [$\%$]\\ 
&$0\leq \mathrm{L_{3D}}\leq20$ [Mpc]&$20\leq \mathrm{L_{3D}}\leq40$ [Mpc]&$40\leq \mathrm{L_{3D}}\leq60$ [Mpc]\\\hline\hline
region 1 &36.0 (6.36)&34.7 (2.67)&39.1 (1.96)\\ 
region 2 &14.6 (4.05)&12.9 (1.63)&7.04 (0.83)\\ 
region 3 &12.4 (3.73)&15.0 (1.75)&19.6 (1.39)\\ 
region 4 &37.1 (6.46)&37.4 (2.77)&34.2 (1.83)\\
\hline
\label{tab.fraction2}
\end{tabular}
\end{center}
\end{table*}
 
\subsubsection{Statistics of haloes}
\label{st.halo}
In Figure~\ref{fig.parameter_reg4}, the averaged values of the halo properties are estimated for different filament mass bins. The values are fitted with a line defined as
\begin{equation}
y=\mathrm{A}~\mathrm{log}\left(x\right)+\mathrm{B},
\label{fittingline}
\end{equation}
where $\mathrm{A}$ and $\mathrm{B}$ are fitting parameters. Fitting is performed to each filament property as a function of filament masses define in equation~(\ref{eq.mass}). The results for fitting are shown in Table~\ref{tab.parameter}. 

\begin{figure*}
\vspace{5cm}
\subfigure{\includegraphics[width=0.92\columnwidth]{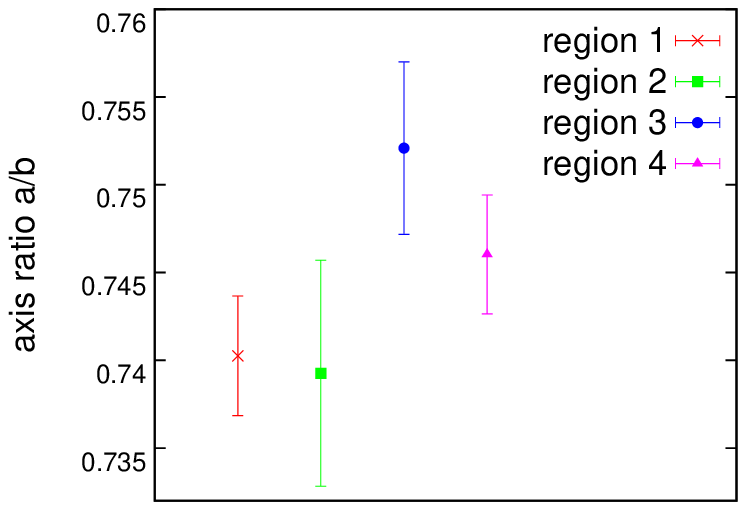}}\hspace{1cm}
\subfigure{\includegraphics[width=0.92\columnwidth]{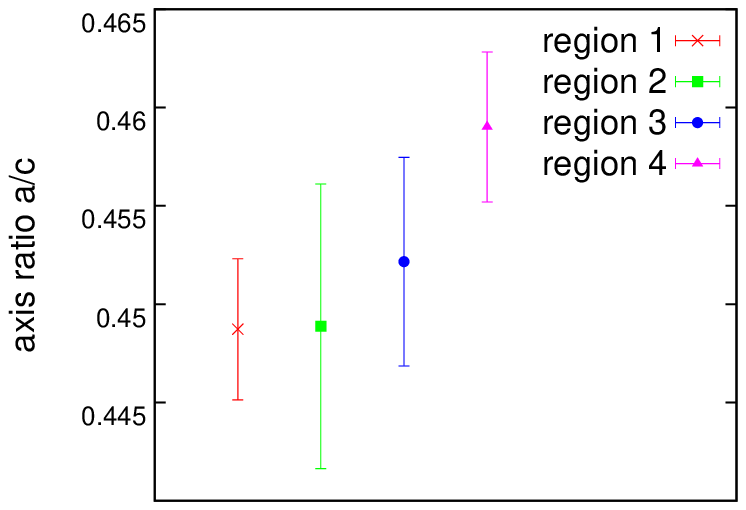}}\hspace{1cm}
\subfigure{\includegraphics[width=0.92\columnwidth]{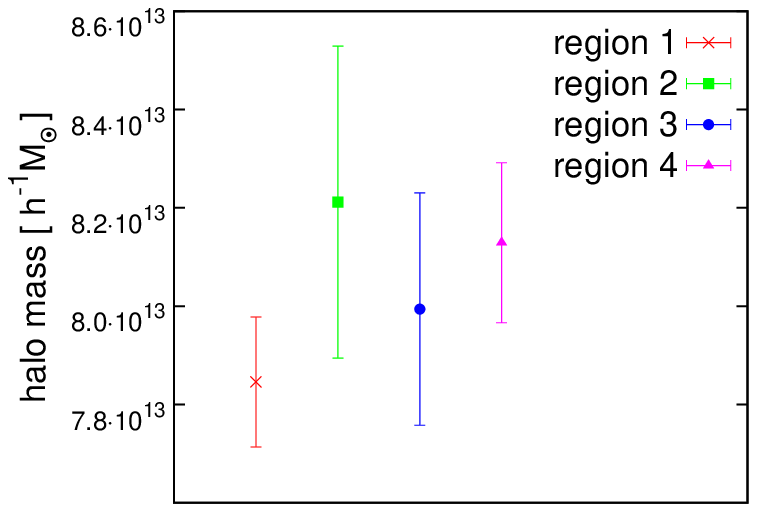}}\hspace{1cm}
\subfigure{\includegraphics[width=0.92\columnwidth]{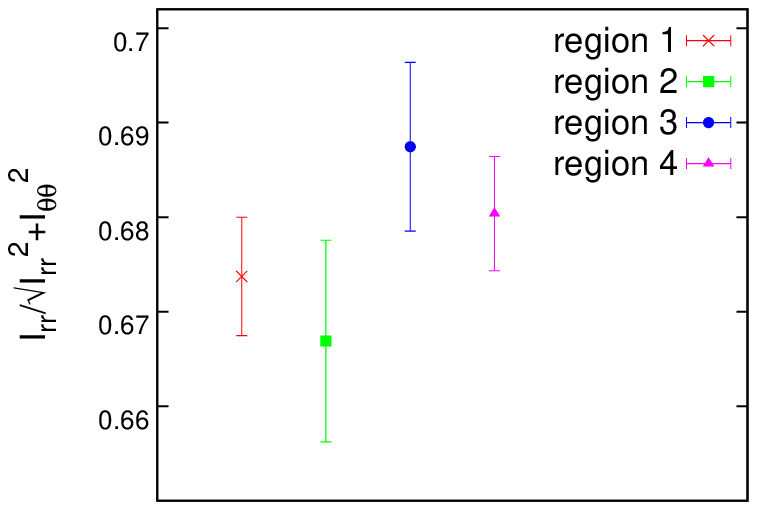}}
\caption{Halo properties for each region. Cross, square, circle and triangle with error bar indicate region~1, region~2, region~3 and region~4, respectively. {\it upper left}: axis ratio a/b; {\it upper right}: axis ratio a/c; {\it lower left}: halo mass $\mathrm{M}_{500}$; {\it lower right}: inertia moment along halo-halo axis $\mathrm{I_{rr}}$.}
\label{fig.parameter_ind}
\end{figure*}

\begin{figure*}
\vspace{5cm}
\subfigure{\includegraphics[width=0.92\columnwidth]{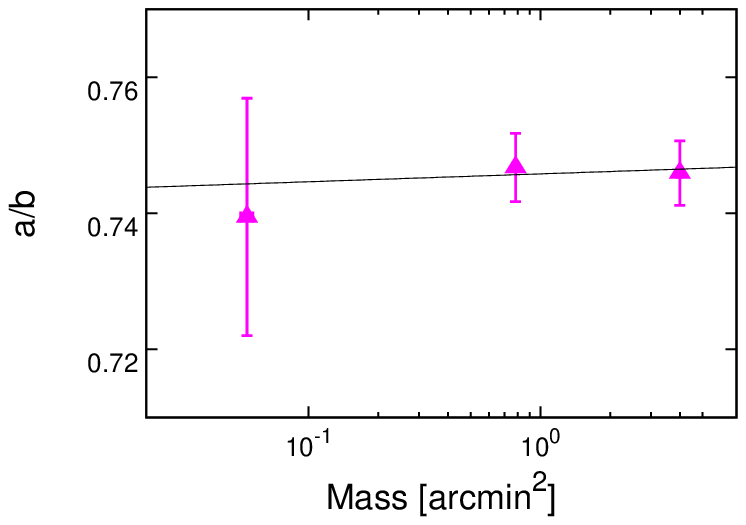}}\hspace{1cm}
\subfigure{\includegraphics[width=0.92\columnwidth]{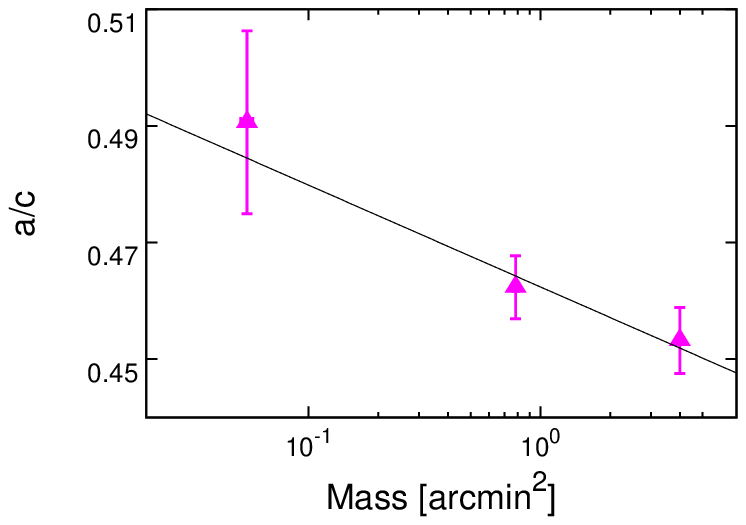}}\hspace{1cm}
\subfigure{\includegraphics[width=0.92\columnwidth]{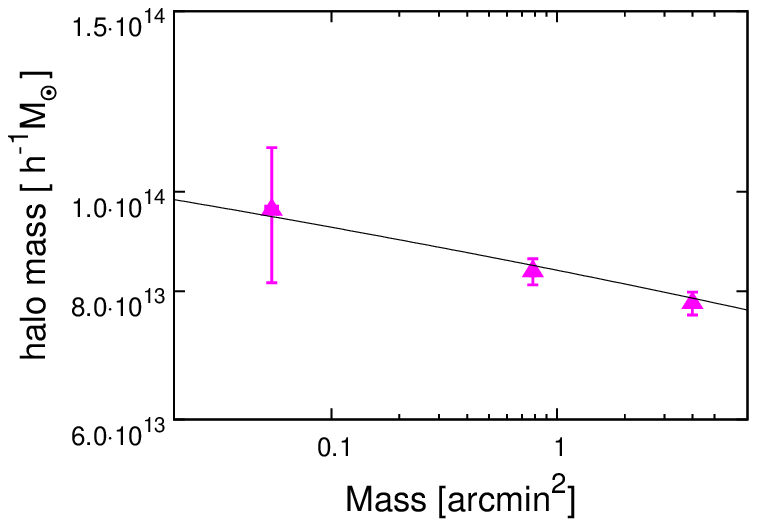}}\hspace{1cm}
\subfigure{\includegraphics[width=0.92\columnwidth]{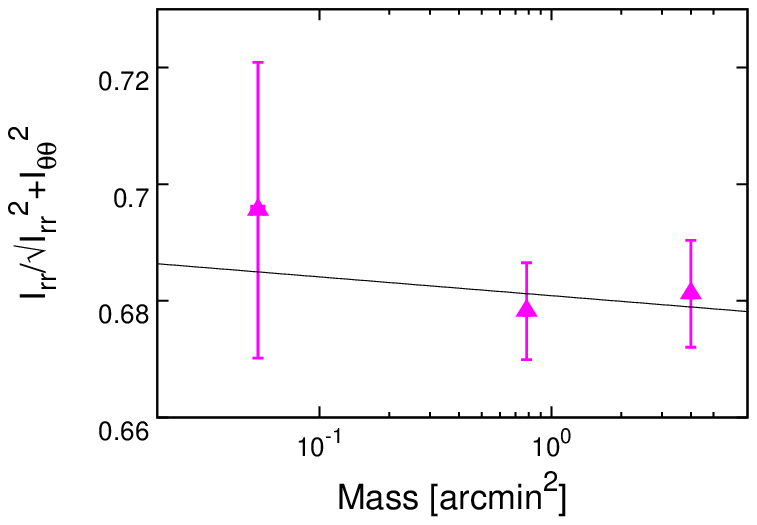}}
\caption{Similar to Figure~\ref{fig.parameter_ind}, but halo properties in region~4 are shown as a function of filament mass. Triangles with error bars show average values in different mass bins. Solid lines show best-fit correlations with equation~(\ref{fittingline}).}
\label{fig.parameter_reg4}
\end{figure*}

 Upper panels in Figure~\ref{fig.parameter_ind} show average axial ratios $\mathrm{a/b}$ and $\mathrm{a/c}$ in each region, which are defined in equation~(\ref{eq.triaxial2}). In region~1 and region 2, we find that the values of both parameters are consistent with each other, because both regions exist under the similar circumstances where other objects are not distributed around their haloes. On the other hand, the values of $\mathrm{a/b}$ in region~3 and region~4 are lager than those in the other regions, and $\mathrm{a/c}$ takes a larger value only in region~4. Generally, we expect that these values in region 3 and 4 are lower than those in region~1 and 2 because haloes in region~3 and 4 are more elongated due to accretion from surrounding structures. While our result therefore appears to be contradict to this naive expectation, our classification method might also affect these statistical properties of haloes. We leave further investigation on this difference as future work.

 Lower left panels in Figure~\ref{fig.parameter_ind} and \ref{fig.parameter_reg4} show the halo mass for each region and only for region~4 as a function of filament mass. In Figure~\ref{fig.parameter_reg4}, the best fit parameters with errors are $(\mathrm{A,B})=(-3.68\pm0.21\times10^{12}, 8.39\pm0.04\times10^{13})$. We find that haloes are less massive with increasing filament masses. This is presumably because massive objects are formed via a sequence of mergers and accretion of smaller objects in the hierarchical structure formation theory. Progenitors of massive haloes reside in filaments, and masses of haloes grow as matter in filaments accrete to these haloes. This suggests that haloes become more massive with decreasing filament masses, as was also implied by \citet{2014arXiv1401.7866C} through $N$-body simulation.
 
 
 Lower right panel in Figure~\ref{fig.parameter_ind} shows the normalized inertia moment along halo-halo axis. 
 We find that the values of the inertia moments along the halo-halo axis in region~3 and 4 are higher than the values in the other regions.
This means that haloes in region~3 and region~4 are elongated along the halo-halo axis due to interaction with filaments. Lower right panel in Figure~\ref{fig.parameter_reg4} shows the same plot only for region~4 as a function of filament mass. The averages for 
different mass bin samples are $\mathrm{I_{rr}}=$($0.680\pm0.023, 0.680\pm0.007$, $0.677\pm0.014$), respectively.

\begin{table*}
\begin{center}
\caption{Fitting results for halo properties. Column (1): halo properties; Column (2) and Column (3) inclination and  intercept of line with $1\sigma$ error.}
\begin{tabular}{ccc}
\hline
&$\mathrm{A}$&$\mathrm{B}$\\ \hline\hline
a/b &$\left(0.50\pm1.21\right)\times10^{-3}$&$\left(7.46\pm0.01\right)\times10^{-1}$\\ 
a/c &$\left(-7.58\pm1.93\right)\times10^{-3}$&$\left(4.62\pm0.02\right)\times10^{-1}$\\ 
halo mass &$\left(-3.68\pm0.32\right)\times10^{12}$&$\left(8.39\pm0.04\right)\times10^{13}$\\
$\mathrm{I_{rr}}$ &$\left(-1.40\pm3.26\right)\times10^{-3}$&$\left(6.81\pm0.04\right)\times10^{-1}$\\ 
\hline
\label{tab.parameter}
\end{tabular}
\end{center}
\end{table*}

\section{Conclusion}
\label{conclusion}
In this paper, we classified dark matter distributions between halo pairs with weak gravitational lensing, and investigated statistical properties of filaments and haloes. We selected filament candidates from the halo catalogue generated from a large set of $N$-body simulations. We classified these candidates into four regions by fitting background-subtracted convergence profiles. We have shown that straight filaments on convergence maps can be classified in the $\kappa_0-\theta_c$ plane, where $\kappa_0$ and $\theta_c$ are two fitting parameters characterizing the convergence profile. 
Specifically, filament candidates were divided into four regions in the $\kappa_0-\theta_c$ plane and filaments in region~4 were found to be straight filaments.  We also found that this classification does not depend on separations between two haloes.
  
We have studied statistical properties of filaments. The number of filaments classified into region~2 and region~4 decreases as a function of separation between the two haloes. On the other hand, the number of filaments in region~1 and region~3 shows increasing trends as separations increase. These trends can be explained by the decrease of correlations with surrounding matter and the projection effect. These trends derived in this paper are broadly consistent with previous study \citep{2005MNRAS.359..272C}. While only $\sim 4$\% of straight filaments in region 4 can be detected with weak lensing, stacked weak lensing allows us to detect the mean mass distribution of filaments easily. In the HSC survey, we can achieve significant detection of filamentary structures at $S/N\geq5$ with stacked lensing.  
The matter density observed with $Planck$ is larger than that used in this paper. This difference would make it easy to form filaments and cause a difference in their statistical properties presented in this paper. Therefore, the statistical property of filaments would serve as a new tool for constraining cosmology.
  
We have also studied statistical properties of haloes at the edges of filaments. 
We found that haloes are less massive with increasing filament masses and are elongated along the halo-halo axis due to interaction with filaments. 
This can be explained by that fact that haloes grow by accretion of matter in filaments. Massive haloes already experienced accretion from filaments, which result in smaller filament masses and more elongation along the halo-halo axis. On the other hand, the dependences of these halo properties on filament masses are very weak, suggesting the necessity of large-scale surveys to observationally confirm these statistical properties.

\section*{Acknowledgements}
We thank an anonymous referee for careful reading and suggestions to improve the quality of the article. We would like to thank Takashi Hamana and Masanori Sato for providing the simulation data. We thank Kohei Hattori for helping the calculation and Yuki Okura for useful discussion. Numerical computations in this paper were in part carried out on the general-purpose PC farm at Center for Computational Astrophysics, CfCA, of National Astronomical Observatory of Japan. This work was supported in part by the FIRST program Subaru Measurements of Images and Redshifts (SuMIRe), World Premier International Research Center Initiative (WPI Initiative), MEXT, Japan, and Grant-in-Aid for Scientific Research from the JSPS (23740161), and in part by Grant-in-Aid for Scientific Research from the JSPS Promotion of Science (23540324, 23740161)
\bibliographystyle{mn2e}
\bibliography{mn-jour,bibtex}
\newpage
\afterpage{\clearpage}

\end{document}